\documentclass[letterpaper,twocolumn,10pt]{article}
\usepackage{usenix,epsfig,endnotes}

\usepackage{balance}
\usepackage{alltt}
\usepackage{amsmath}
\usepackage{balance}
\usepackage{booktabs}
\usepackage{fixltx2e}
\usepackage{graphicx}
\usepackage{boxedminipage}
\usepackage{hyperref}
\usepackage{nicefrac}
\usepackage{subfig}
\usepackage{setspace}
\usepackage{xspace}
\usepackage{multirow}
\usepackage{colortbl}
\usepackage{amsfonts} 
\usepackage{blindtext}
\usepackage{chngpage}
\usepackage{listings}
\usepackage{color}
\usepackage[dvipsnames]{xcolor}
\usepackage{mathtools}
\usepackage{amssymb}
\usepackage{pifont}
\usepackage[numbers,sort&compress,square]{natbib}

\usepackage{wrapfig}

\RequirePackage{listings}

\lstdefinelanguage{Golang}%
  {morekeywords=[1]{package,import,func,type,struct,return,defer,panic,%
     recover,select,var,const,iota,},%
   morekeywords=[2]{string,uint,uint8,uint16,uint32,uint64,int,int8,int16,%
     int32,int64,bool,float32,float64,complex64,complex128,byte,rune,uintptr,%
     error,interface},%
   morekeywords=[3]{map,slice,make,new,nil,len,cap,copy,close,true,false,%
     delete,append,real,imag,complex,chan,},%
   morekeywords=[4]{for,break,continue,range,goto,switch,case,fallthrough,if,%
     else,default,},%
   morekeywords=[5]{Println,Printf,Error,Print,},%
   sensitive=true,%
   morecomment=[l]{//},%
   morecomment=[s]{/*}{*/},%
   morestring=[b]',%
   morestring=[b]",%
   morestring=[s]{`}{`},%
   }

\definecolor{dkgreen}{rgb}{0,0.6,0}
\definecolor{gray}{rgb}{0.5,0.5,0.5}
\definecolor{mauve}{rgb}{0.58,0,0.82}

\newcommand{\beforecaption}{\vspace{-.15cm}\begin{spacing}{0.85}}
\newcommand{\aftercaption}{\vspace{-.45cm}\end{spacing}}

\begin{document}

\twocolumn[
\begin{@twocolumnfalse}
\begin{center}
{\Large\bf Fine-Grained Library Customization}
\end{center}
\smallskip
\vspace{-0.2in}
\begin{center}
Linhai Song,
Xinyu Xing \\
Pennsylvania State University
\end{center}

\bigskip
\end{@twocolumnfalse}
]



\begin{abstract}
Code bloat widely exists in production-run software. Left untackled, it not only
degrades software performance but also increases its attack surface. In this
work, we conduct a case study to understand this issue in statically linked
libraries. To be specific, we analyze \texttt{midilib}, a software package
enclosing statically linked libraries. We show that it is possible to leverage
dependence analysis to trim the resultless code statements residing in a
target library. With this observation, we believe it is possible to build a tool
to automatically cut off code pertaining to resultless operations.

\end{abstract}

\section{Introduction}
\label{sec:intro}

Modular design is widely used in traditional software development  to control
the implementation complexity~\cite{modular}. After dividing a large program
into smaller modules or libraries, developers can focus on their own parts and
implement desired functionalities.  To improve development productivity,
libraries are encouraged to be reusable  and to be shared by different
programs~\cite{library}.  Therefore, libraries tend to have generic interfaces
and provide  different functionalities for various usage scenarios.

When a library is used by a program, it is usually statically linked to that
program. Since the program only has limited calling contexts and usage
scenarios,  more-than-necessary code inside the library is linked,  causing code
bloat~\cite{code-bloat}. Bloated code widely exists in production-run software.
For example, a recent study shows that only around 20\%  instructions of Firefox
are executed under  typical workloads~\cite{code-bloat-study}.

Bloated code can lead to various problems.  First, it potentially introduces
more bugs and vulnerabilities. A recent study has showed that most
vulnerabilities in protocol implementation  reside in modules not widely
used~\cite{protocol-mao}.  Second, a larger code size increases memory pressure
and causes cache misses when loading instructions~\cite{PBI}. Third, bloated
code incurs resultless or redundant computation,  resulting in computation
inefficiency~\cite{BloatFSE2008,XuBloatPLDI2009,XuBloatPLDI2010}. Last but not
least, a larger code size also consumes network bandwidth  when being
distributed across the Internet~\cite{container-debloating-1, 
container-debloating-2}.

Inspired by this, we propose to address the code bloating problem by performing
library customization against  statically linked libraries. Different from
previous works on detecting runtime bloat~\cite{BloatFSE2008,XuBloatPLDI2009,
XuBloatPLDI2010,PerfBlower,Reusable,Cachetor}  or library
customization~\cite{ldoctor, protocol-mao, container-debloating-1, 
container-debloating-2, dinghao-1, dinghao-solo-1, dinghao-solo-2}, our technique 
is a fine-grained code removal scheme built on the basis of the following
hypothesis. Many library functions  return its computation results as a data
object defined by a ``struct''. However, it is typical the case that many fields
in the struct are not in use by the upper level applications. This means that
there must be resultless computation residing in the library and we have the
potential to trim the code pertaining to the resultless computation.

To validate our hypothesis, we conduct a case study against a software package
which contains statically linked libraries. We show that a library returns a data
object with 44 primitive fields. However, upper level software only uses 9 of
them. By using dependence analysis along with two code trimming schemes, we
can reduce the code space of the library by about 50\%. Given that many software 
contain statically linked libraries, we believe this observation and practice could be 
potentially generalized and significantly benefit library customization.



The rest of the paper is organized as follows. 
In Section~\ref{sec:case}, we discuss our case study against a software package and describe two proof-of-concept techniques. 
In Section~\ref{sec:related}, we discuss the works relevant to library customization. In Section~\ref{sec:conclusion},  
we conclude our paper and discuss future works.

\begin{table}[h]
\begin{tabular}{c}
\hspace{12pt}

\begin{lstlisting}  
/* m2rtttl.c */
while(midiReadGetNextMessage(mf, &msg)) {
  ...
  switch(msg.iType) {
    case msgNoteOff:
      if(iChannel==msg.MsgData.NoteOff.iChannel) {
        if(iCurrPlayingNote==msg.MsgData.NoteOff.iNote) {
          outStdout(...);
          iCurrPlayingNote = -1;
          iCurrPlayStart = msg.dwAbsPos;
        }
      }
      break;
    case msgNoteOn: ... 
      break;
    case msgMetaEvent: ...
      break;
    default: /* Ignore other cases */
      break;
  }
}
\end{lstlisting}

\end{tabular}
\caption{The code fragment of the \texttt{while} loop implementing the 
major functionality of \texttt{m2rtttl}.}
\label{code:m2rtttl}
\end{table}

\section{Case Study}
\label{sec:case}

As is mentioned above, a shared library statically linked might contain many
resultless operations that can be potentially trimmed. In this section, we
illustrate this practice by taking for example midilib~\cite{midilib} an 
open-source repository in C.


Midilib contains an implementation of I/O libraries for  MIDI files~\cite{midi}
in \texttt{midifile.c}. It also provides other functionalities. After building
midilib, we can obtain five executables,  which are \ding{182} \texttt{m2rtttl},
to convert MIDI files to RTTTL~\cite{rtttl},  \ding{183} \texttt{mididump}, to
dump MIDI file content, \ding{184} \texttt{mfc120}, to change MIDI file version,
\ding{185} \texttt{mozart}, to generate simple musics, and \ding{186}
\texttt{miditest}, to conduct tests. The compiled object file of
\texttt{midifile.c} is statically linked to every executable. Thus, each
executable contains the same library.

In the following, we first perform simple code analysis against the
aforementioned software and thus reveal those resultless operations residing in
the shared library. More specifically, we perform our analysis on the executable
\texttt{m2rtttl} as well as the library implemented in \texttt{midifile.c}.
Second, we introduce a prototype system to demonstrate the potential of
debloating the library. It should be noted that we do not claim our prototype
system is an effective solution for dealing with the library debloating. Rather,
it is just a preliminary proof-of-concept tool, demonstrating the possibility
of reducing the code space for the library.


\begin{table}[t]
\begin{tabular}{c}
\hspace{12pt}

\begin{lstlisting}  
typedef struct {
  tMIDI_MSG iType;
  DWORD   dt;
  DWORD   dwAbsPos;
  ...
  union {
    struct {
      int iNote;
      int iChannel;
      int iVolume;
    } NoteOn;
    struct {
      int iNote;
      int iChannel;
    } NoteOff;
    struct { ... } NoteKeyPressure;
    struct { ... } NoteParameter;
    ...
  } MsgData;
  ...
} MIDI_MSG;
\end{lstlisting}

\end{tabular}
\caption{The code fragment of struct \texttt{MIDI\_MSG}.}
\label{code:struct}
\end{table} 

\subsection{Code Analysis and Key Observation}
\label{subsec:obs}
 
The major functionality of \texttt{m2rtttl} is implemented using  a
\texttt{while} loop shown in Table~\ref{code:m2rtttl}. In each iteration, the
\texttt{while} loop reads a midi packet from an input  file on line 2 and
changes the packet to a string in RTTTL format on line 8. The function
\texttt{midiReadGetNextMessage} is implemented in the shared library
\texttt{midifile.c}. It parses an input file, indicated by its argument
\texttt{mf},  and returns midi packets through argument \texttt{msg}. In
\texttt{midilib} repository, \texttt{msg} is in \texttt{MIDI\_MSG} struct type.
As is shown in Table~\ref{code:struct}, \texttt{MIDI\_MSG} contains an
\texttt{enum} field \texttt{iType},  indicating the type of a midi packet. In
addition, \texttt{MIDI\_MSG} contains a \texttt{union} field \texttt{MsgData} to
hold data for different types of midi packets.

Inside the \texttt{while} loop depicted in Table~\ref{code:m2rtttl}, we
enumerate the fields in struct \texttt{MIDI\_MSG} and count the number of
accesses to those fields. Based on the software implementation, we observe that
\texttt{MIDI\_MSG} defines 44 primitive fields, among which the \texttt{while}
loop accesses only 9 fields. Intuition suggests,  if there is no access to a
field in the upper level of applications, the statements in the library tied to
that field might represent a set of resultless operations. As a
result, we believe that we might be able to trim the implementation of the
shared library \texttt{midifile.c} by carefully cutting off those statements
pertaining to the resultless operations.

In addition to the field enumeration and the examination of field accesses, we
count how many packet types can be processed by the \texttt{while} loop and how
many packet types can be generated by \texttt{midiReadGetNextMessage}. We
observe that the library function \texttt{midiReadGetNextMessage} assembles 9
different types of midi packets, whereas the \texttt{while} loop only processes
3 types (i.e., \texttt{msgNoteOn}, \texttt{msgNoteOff} and
\texttt{msgMetaEvent})  and simply ignores other types on line 18 in
Table~\ref{code:m2rtttl}. Intuition suggests that we might be able to leverage
those unused packet types as another indicator to identify those resultless
operations in the lower level library. This is because the \texttt{while} loop
does not impose any computation upon 6 types of packets, and when computation
pertaining to  those types of packets present in the library, they represent the
operations irrelevant and futile.

Last but not least, inside the code base in the library, we measure the amount
of read and write operations that tie to the fields not being accessed by the
upper level of the application (i.e., \texttt{m2rtttl}). By using \texttt{LLVM}
to instrument the library \texttt{midifile.c} and performing 10 runs 
with different inputs, in total, we track down 4755 read and 2833 write operations 
pertaining to the fields of the struct \texttt{msg}. 
Among the 2833 write operations, we note 
that 1133 operations cannot be removed because the fields tied to these write operations
have been read by the upper level of the application. For the remaining
1700 write operations, we partition them into two categories. In the first
category, we observe that, there are 1015 write operations (about 60\%) that
neither the lower level of the library nor the upper level of the application
read the fields tied to these operations, which implies that we can
safely trim these operations and thus reduce the code space of the library. In
the second category indicated by the remaining 685 write operations
(approximately 40\%), we note that, while the fields tied to these operations
are not read by the upper level of the application, they are involved in
data dependency in lower level of the library. Admittedly, this does not mean
we cannot remove these operations. But it implies that we have to trim these
operations with the consideration of data dependency.

\begin{table}[t]
\begin{tabular}{c}
\hspace{12pt}

\begin{lstlisting}  
/* midifile.c */
BOOL midiReadGetNextMessage(..., MIDI_MSG *msg) {
  ...
  switch(msg.iType) {
    case msgNoteOn: 
      msg->MsgData.NoteOn.iChannel = ...;
      msg->MsgData.NoteOn.iNote = ...;
      msg->iMsgSize = 3;
      break;
    case msgNoteOff: ... 
      break;
    case msgNoteKeyPress: 
      msg->MsgData.NoteKeyPress.iChannel = ...;
      msg->MsgData.NoteKeyPress.iNote = ...;
      msg->MsgData.NoteKeyPress.iPressure = ...;
      msg->iMsgSize = 3;
      break;
    case msgSetParameter: ...
      break;
    case msgSetProgram: ...
      break;
    case msgChangePress: ...
      break;
    case msgSetPitchWheel: ...
      break;
    case msgMetaEvent: ...
      break;
    case msgSysEx1:
    case msgSysEx2: ...
      break;
  }
  ptr += msg->iMsgSize
  ...
}
\end{lstlisting}

\end{tabular}
\caption{The code fragment of the library function \texttt{midiReadGetNextMessage}.}
\label{code:readnext}
\end{table} 

\subsection{Customization Demonstration}
\label{subsec:cust}

Based on the analysis and observation above, we design and develop two simple
tools using \texttt{LLVM}  to customize the lower level of the library or more
precisely speaking the function \texttt{midiReadGetNextMessage} implemented in
the library. In the following, we describe them in turn.



\noindent\textbf{Tool for Eliminating Resultless Field Assignments.}
As is mentioned above, if the value assigned to a field is not accessed by
the upper level of the application nor the lower level of the library, then we
could safely remove the statements tied to such assignment. Inspired by this
observation, we develop a tool to identify and cut off such code statements.

To be specific, we first leverage the struct layout information provided by
\texttt{LLVM} to compute the offsets of the fields that neither the library nor
the application reads. Then, using this information, we pinpoint those
assignment instructions (i.e., \texttt{LLVM} intermediate code) corresponding to
these fields. Since the assignment instructions represent the site of assigning
a value to a field, and operations pertaining to such sites also contain those 
instructions that compute the assigned value and the field address,
we finally perform a simple data dependence analysis to further identify the instructions
relevant to the field assignment. In this preliminary work, we deem such
instructions as unnecessary and our tool trim these instructions along with
those tied to resultless value assignments.

In the implementation of the library function \texttt{midiReadGetNextMessage},
our tool tracks down 4 fields, the read of which neither presents in the
library nor the upper level of the application. These 4 fields associate with 51
instructions in the library, accounting for about 7\% (51 out of 722)
\texttt{LLVM} intermediate code that we can safely eliminate. It is not
difficult to note, as is mentioned in Section~\ref{subsec:obs}, we have
identified only 9 fields that the upper level of the application reads,
but our tool tracks down only 4 fields, the read of which are not
presented in the application. Here, the reason is as follows.

As is illustrated in Figure~\ref{code:struct}, many primitive fields are
enclosed in the \texttt{union} type field \texttt{MsgData}. From the perspective
of \texttt{LLVM}, this means that the machine utilizes the same memory location
to store different struct fields, such as \texttt{NoteOff}, \texttt{NoteOn},
\texttt{NoteKeyPressure}, and so on. In our implementation, our tool
distinguishes primitive fields based on their offsets. This means it lacks the
ability to distinguish the primitive fields referred by the union struct, 
such as failing to differ \texttt{msg.MsgData.NoteOn.iNote} from 
\texttt{msg.MsgData.NoteOff.iNote}.
Therefore, our current results miss to pinpoint some primitive fields 
that neither the library
nor the application reads.

\noindent\textbf{Tool for Eliminating Unused Packet Types.}
Recall that in addition to leveraging resultless field assignments for library code
customization, we can use the packet type information as an indicator to identify
resultless operations in the library. Motivated by this observation, we design
and develop another tool that takes advantage of this observation and performs
library code customization.

As is shown in Table~\ref{code:readnext}, the library assigns a value to a field
(e.g., \texttt{msg.MsgData.NoteKeyPress.iChannel} in line 13) only after it
reads another field (e.g., \texttt{msg.iType} in line 4) and a certain
condition holds (e.g., \texttt{msg.iType} == \texttt{msgNoteKeyPress} in line
5). Intuition suggests this can be interpreted as a control dependency.
In the higher level of the application, as is shown in Table~\ref{code:m2rtttl},
we do not observe the same control dependency relationship. This means
that the corresponding computation in the library (i.e., the statements in line
13-15) has no effect upon the upper level of the application. We could leverage
this mismatched pattern to cut off the code fragment accordingly and thus reduce
the code size of the library. It should be noted that we do not trim the
statements in line 16-17 nor that in line 12 depicted in
Table~\ref{code:readnext} because -- as is specified in line 32 -- the global
variable \texttt{ptr} is dependent upon \texttt{iMsgSize}.

In this work, we implement this pattern matching approach using \texttt{LLVM}.
By performing code customization against the aforementioned library through the
patterns identified, we track down 33 field assignment instructions in the
library that do not have impact to the upper level of the application. Following
the procedure used in the first tool mentioned above, we also use data
dependence analysis to pinpoint other statements pertaining to those tied to
resultless filed assignments. Together with the 33 resultless instructions, in
total, we pinpoint 355 out of 722 \texttt{LLVM} instructions that can be safely
eliminated.

Going beyond testing the tools individually, we further combine the unnecessary
instructions obtained from both aforementioned tools. We observe the combined
techniques can identify 36 resultless field assignments in total. Using  data
dependence analysis, they lead up to the removal of 367 \texttt{LLVM}
instructions, accounting for 50.83\% of instructions removal (i.e., 367 out of 722).
We have already noted that these about 50\% of instructions removal reflect
approximately 39.63\% lines of source code removal. This indicates the potential
of fine-grained library customization.

\section{Related Works}
\label{sec:related}

Code bloat~\cite{code-bloat} refers to unnecessarily large code size, 
which can increase security attack surface, consume more memory,
lower instruction cache performance, and even make the distribution of software more difficult. 
There are empirical studies that confirm the existence of code bloat and its negative impact.
\citet{code-bloat-study} conduct an empirical study to understand how much unused code
in different types of programs. 
The authors propose two methods to measure the size of unused code, 
one is to identify function calls isolated from call graph statically, 
and the other is to dynamically profile how many instructions are not executed under typical workloads. 
The authors report that a large portion of code is not 
executed in the investigated programs.
\citet{protocol-mao} study 20 vulnerabilities related to protocol implementation,
and finds that most of the vulnerabilities reside in code implementation not commonly used.
Customizing protocol implementation can successfully eliminates most of these vulnerabilities.

Researchers and practitioners build many techniques to identify and eliminate unnecessary 
functionalities~\cite{ldoctor, protocol-mao, container-debloating-1, container-debloating-2,dinghao-1, dinghao-solo-1, dinghao-solo-2}.
LDoctor~\cite{ldoctor} identifies inefficient loops conducting resultless computation 
and suggests developers remove these loops conditionally or unconditionally.  
\citet{protocol-mao} proposes a feature access control system to unify protocol implementation customization, 
which can remove unnecessary features in protocol implementation and reduce attack surface. 
Application containers usually contain unneeded files. 
As a dynamic technique,
Cimplifier~\cite{container-debloating-1, container-debloating-2} can automatically detect 
unnecessary resources through analyzing system calls. 
JRed~\cite{dinghao-1} detect unused classes and methods using reachability analysis after building call graphs for programs to be customized. 
\citet{dinghao-solo-1,dinghao-solo-2} propose a technique to cut user-specified functions through backward and forward slicing. 
Although useful, these techniques work on code granularities 
much larger than our proposed techniques, such as loops or files.
These techniques do not target to eliminate fine-grained resultless computation.

\section{Conclusion and Future Work}
\label{sec:conclusion}

In this paper, we perform a code analysis against a library statically linked to
a target executable. We show that using the fields in a data structure we can
potentially trim the code statements resultless for the upper level of
applications, and thus potentially reduce the attack surface of the library.
Using two proof-of-concept tools designed and developed based on our analysis,
we demonstrate the possibility of performing fine-grained customization for a
library. 

As future work, we will extend our techniques from the following aspects.
First, we plan to examine more libraries and conduct an empirical study to 
understand root causes of resultless field assignments and their impact in the real world.
Second, we plan to build a robust static technique 
and explore different design points during technical design.
Third, we plan to build an automated testing platform 
combining static and dynamic analysis to test customized libraries.


\balance
{
\bibliographystyle{abbrvnat}
\bibliography{protocol} 
}
\end{document}